\documentclass[aps,showpacs,superscriptaddress]{revtex4}
\usepackage{graphicx}
\usepackage{amsmath}
\usepackage{bm}
\begin{document}
\title{Singular Energy Distributions in Granular Media}
\author{E.~Ben-Naim}
\email{ebn@lanl.gov}
\affiliation{Theoretical Division and Center for Nonlinear Studies,
Los Alamos National Laboratory, Los Alamos, New Mexico 87545 USA}
\author{A.~Zippelius}
\email{annette@theorie.physik.uni-goettingen.de}
\affiliation{Institut f\"ur Theoretische Physik,
Georg-August-Universit\"at, 37077 G\"ottingen, Germany}
\begin{abstract}
  We study the kinetic theory of driven granular gases, taking into
  account both translational and rotational degrees of freedom. We
  obtain the high-energy tail of the stationary bivariate energy
  distribution, depending on the total energy $E$ and the ratio
  $x=\sqrt{E_w/E}$ of rotational energy $E_w$ to total energy.
  Extremely energetic particles have a unique and well-defined
  distribution $f(x)$ which has several remarkable features: $x$ is
  not uniformly distributed as in molecular gases; $f(x)$ is not
  smooth but has multiple singularities. The latter behavior is
  sensitive to material properties such as the collision parameters,
  the moment of inertia and the collision rate. Interestingly, there
  are preferred ratios of rotational-to-total energy.  In general,
  $f(x)$ is strongly correlated with energy and the deviations from a
  uniform distribution grow with energy. We also solve for the energy
  distribution of freely cooling Maxwell Molecules and find
  qualitatively similar behavior.
\end{abstract}
\maketitle
\section{Introduction}

Energy dissipation has profound consequences in granular materials,
especially in dilute gases, where the dynamics are controlled by
collisions \cite{bp,pl,pb}. Dissipation is responsible for many
interesting collective phenomena including clustering
\cite{my,ou,lh,nbc,vvl}, formation of shocks
\cite{bcdr,elm,zbh,rbss,smk}, and hydrodynamic instabilities
\cite{gz,km}. Another consequence is the anomalous statistical physics
that includes the non-Maxwellian velocity distributions
\cite{lcdkg,rm,ao,dlk,kwg,gzb} and the breakdown of energy
equipartition in mixtures \cite{wp,fm}.

For an elastic gas in equilibrium, the temperature, defined as the
average kinetic energy, characterizes the entire distribution function
including all of the moments, the bulk of the distribution, as well as
the tail of the distribution. Outside of equilibrium, the temperature
is not sufficient to characterize the energy distribution.  Granular
gases are inherently out of equilibrium and a complete
characterization must therefore include the behavior of typical
particles, the behavior of energetic particles, as well as the moments
of the distribution. For example, the energy distribution may have
power-law tails with divergent high-order moments \cite{kb,eb,bmp} and
consequently, the moments exhibit multiscaling \cite{bk}. Generally,
nonequilibrium effects are pronounced in the absence of energy input
to balance the dissipation but can be suppressed by injection of
energy where the deviation from a Maxwellian distribution affects only
extremely energetic particles \cite{rm,ve,bk1,kssaob}.

While there is substantial understanding of the energy distribution of
frictionless granular gases, much less is known theoretically
\cite{sz,tg,hz,lhnz,hhz,jz,hczhl,hob} and experimentally
\cite{Walton,flca,trgl} when the rotational degrees of freedom are taken
into account. It is difficult to measure the rotational motion
experimentally, and the few available measurements are restricted to
two-dimensions. Surface roughness and friction have important
consequences and the hydrodynamic theory \cite{jr,bdks,bdl,ig} must be
modified, if the particles have spin \cite{gnb}.  Equipartition does
not hold for the average rotational and translational temperature --
neither in the free cooling case \cite{hz,lhnz,hhz,jz} nor for a
driven system \cite{hczhl}.  In general, rotational and linear degrees
of freedom are correlated in direction \cite{bpkz}.

In this paper, we investigate the nature of the full energy
distribution, that is, the bivariate distribution of rotational and
translational energy. Motivated by the fact that on average the total
energy is not partitioned equally between rotational and translational
degrees of freedom, we focus on the bivariate distribution $P(E,x)$ of
total energy $E$ and the modified ratio $x=\sqrt{E_w/E}$ of rotational
to total energy. We thereby generalize the understanding of
frictionless granular matter in terms of the energy distribution to
rough grains.

Our starting point is the nonlinear Boltzmann equation with a
collision rule that accounts for the coupling of translational and
rotational motion due to tangential restitution.  We study stationary
solutions of the inelastic Boltzmann equation that describe steady
states achieved through a balance between energy injections that are
powerful but rare and energy dissipation through inelastic
collisions. For high-energy particles we derive a linear equation for
the bivariate energy distribution. The latter can be shown to
factorize -- $P(E,x)=p(E)f(x)$ -- into a product of the distribution
of the total energy, $p(E)$, and the distribution of the fraction of
energy stored in the rotational degrees of freedom, $f(x)$. The former
distribution decays algebraically with energy: $p(E)\sim
E^{-\nu}$. The fraction of energy stored in rotational motion is
universal for energetic particles in the sense that $f(x)$ approaches
a limiting distribution independent of energy.  Furthermore, this
quantity has a number of interesting features.  First, the
distribution is not uniform, as it would be, if equipartition were to
hold. Second, the distribution is not analytic but has singularities
at special energy ratios. Third, the distribution and in particular
its singularities depend sensitively on the moment of inertia and the
collision parameters. Only for energetic particles is this
distribution well defined. In general, the partition of energy into
rotational and translational motion depends on the magnitude of the
energy.  This paper specifically addresses two-dimensions, although
the theoretical approach and the reported qualitative behavior are
generic.

We also develop a general framework for describing high-energy
collisions and we use this framework to study freely cooling Maxwell
Molecules where the moments of the energy distribution can be found in
a closed form.  For example, the two granular temperatures
corresponding to the rotational and translational motions are coupled
and generally, they are not equal. The high-energy behavior found for
driven steady-states extends to freely cooling gases.

The rest of this paper is organized as follows. We review the
collision rules and introduce the nonlinear kinetic theory in section
II. We then derive the linear kinetic theory for high-energy particles
in section III. Next, in section IV, we study driven steady states and
solve for the stationary energy distribution.  Freely cooling Maxwell
molecules are discussed in section V and we conclude in section
VI. The Appendices detail technical derivations.

\section{The Nonlinear Kinetic Theory}

Our system consists of an infinite number of identical particles with
mass $m=1$, radius $R$, and moment of inertia $I=qR^2$ where $0\leq
q\leq 1$ is a dimensionless quantity. Each particle has a linear
velocity ${\bf v}$ and an angular velocity ${\bf w}$. Its total energy
is shared by the linear and the rotational motion, $E=E_v+E_w$, or explicitly,
\begin{equation}
\label{e-def}
E=\frac{1}{2}\left(v^2+qR^2w^2\right)
\end{equation}
where $v\equiv |{\bf v}|$ and $w \equiv |{\bf w}|$.

In a collision between two particles, their velocities $({\bf
v}_i,{\bf w}_i)$ with the labels $i=a,b$, change according to
\begin{equation}
\label{collision} ({\bf v}_a,{\bf w}_a)+({\bf v}_b,{\bf w}_b)\to\
({\bf v}'_a,{\bf w}'_a)+({\bf v}'_b,{\bf w}'_b)
\end{equation}
where the postcollision velocities are denoted by primes.  In a binary
collision, rotational and translational energy are exchanged, while
the total energy decreases.  In this study, we consider tangential
restitution in addition to the standard normal restitution. Let ${\bf
r}_i$ be the position of particle $i$, then the directed unit vector
connecting the centers of the colliding particles is \hbox{$\hat{\bf
n}=({\bf r}_a-{\bf r}_b)/|{\bf r}_a-{\bf r}_b|$}.  We term this vector
the impact direction. The collision rules are most transparent in
terms of ${\bf u}_i$ the particle velocity at the contact point
\begin{subequations}
\label{relative}
\begin{align}
{\bf u}_a&={\bf v}_a+R\,\hat{\bf n}\times{\bf w}_a\\
{\bf u}_b&={\bf v}_b-R\,\hat{\bf n}\times{\bf w}_b.
\end{align}
\end{subequations}
The inelastic collision laws state that the normal component of the
relative velocity ${\bf U}={\bf u}_a-{\bf u}_b$ is reversed and
reduced by the normal restitution coefficient $0\leq r_n\leq 1$. The
tangential component is either reversed (rough particles) or not
(smooth particles) and in any case reduced by the tangential
restitution coefficient $-1\leq r_t\leq 1$, according to the following
collision rules:
\begin{subequations}
\label{law}
\begin{align}
{\bf U}'\cdot \hat{\bf n}&=
-r_n{\bf U}\cdot \hat{\bf n},\\
{\bf U}'\times \hat{\bf n}&=
-r_t\,{\bf U}\times \hat{\bf n}.
\end{align}
\end{subequations}

Inelastic collisions conserve linear and angular momentum.
Conservation of linear momentum implies that the total linear
velocity does not change, and conservation of angular momentum
enforces that the angular momentum of each particle with respect to the
point of contact remains the same, because there is no torque acting at
the point of contact. The collision laws \eqref{law} combined with these
conservation laws specify the postcollision velocities as linear
combinations of the precollision velocities \cite{hz}
\begin{subequations}
\label{rule}
\begin{align}
{\bf v}_a'&={\bf v}_a-\eta_n{\bf V}\cdot \hat{\bf
  n}\,\hat{\bf n} -\eta_t\left({\bf V}-{\bf V}\cdot \hat{\bf
  n}\,\hat{\bf n}\right)
-\eta_t R \,\hat{\bf n}\times {\bf W}\qquad {\bf w}'_a
={\bf w}_a+\frac{\eta_t}{qR}\,\hat{\bf n}\times {\bf V}+
\frac{\eta_t}{q} \,\hat{\bf
  n}\times\hat{\bf n}\times{\bf W}\\
{\bf v}_b'&={\bf
  v}_b+\eta_n{\bf V}\cdot \hat{\bf
  n}\,\hat{\bf n} +\eta_t\left({\bf V}-{\bf V}\cdot \hat{\bf
  n}\,\hat{\bf n}\right)
+\eta_t R \,\hat{\bf n}\times {\bf W}\qquad\,
{\bf w}'_b=
{\bf w}_b+\frac{\eta_t}{qR}\,\hat{\bf n}\times {\bf V}+ \frac{\eta_t}{q}
\,\hat{\bf
  n}\times\hat{\bf n}\times {\bf W}
\end{align}
\end{subequations}
where the shorthand notations \hbox{${\bf V}={\bf v}_a-{\bf v}_b$} and
\hbox{${\bf W}={\bf w}_a+{\bf w}_b$} were introduced. These collision
rules involve the normal and tangential collision parameters, defined
as
\begin{equation}
\label{eta} \eta_n=\frac{1+r_n}{2}, \qquad {\rm and}\qquad
\eta_t=\frac{q}{1+q}\frac{1+r_t}{2}.
\end{equation}
Their range of values is bounded by \hbox{$1/2\leq \eta_n\leq 1$}
and \hbox{$0\leq \eta_t\leq q/(1+q)$}.  Details of the derivation of
the collision rules are given in Appendix A, as they are relevant
for our discussion. The energy dissipation, $\Delta
E=E_a+E_b-E_a'-E_b'$, is given by
\begin{eqnarray}
\label{dissipation} \Delta E= \frac{1-r_n^2}{4}({\bf V}\cdot\hat{\bf
n})^2+ \frac{q}{1+q}\frac{1-r_t^2}{4}({\bf V}-{\bf V}\cdot\hat{\bf
n}\,\hat{\bf n}+ R\,\hat{\bf n}\times{\bf W})^2.
\end{eqnarray}
The energy dissipation is always positive, except when the collisions
are elastic, $r_n=1$ and $r_t=-1$ (perfectly smooth spheres) or
$r_t=1$ (perfectly rough spheres).

The collision rate $K({\bf v}_a,{\bf v}_b)$ is the rate by which the
two particles approach each other. For hard spheres, this rate is
simply the normal component of the relative velocity, but we study the
general case
\begin{equation}
\label{rate}
K({\bf v}_a,{\bf v}_b)=|({\bf v}_a-{\bf v}_b)\cdot \hat{\bf n}|^\gamma
\end{equation}
with $0\leq\gamma\leq 1$.  Of course, the collision rate vanishes,
$K=0$, when the particles are moving away from each other, $({\bf
v}_a-{\bf v}_b)\cdot \hat{\bf n}>0$. When particles interact via the
central potential $r^{-\kappa}$ then $\gamma=1-2\frac{d-1}{\kappa}$
\cite{rd}. The two limiting cases are hard spheres ($\gamma=1$) and
Maxwell molecules ($\gamma=0$) where the collision rate is
independent of the velocity \cite{jcm,krupp,tm,mhe}.

The central quantity in kinetic theory is the probability $P({\bf v},{\bf
w},t)$ that a particle has the velocities $({\bf v},{\bf w})$ at
time $t$. We study spatially homogeneous situations where this
velocity distribution function is independent of position.  Under
the strong assumption that the velocities of the two colliding
particles are completely uncorrelated, the velocity distribution
obeys the Boltzmann equation
\begin{eqnarray}
\label{nonlinear}
\frac{\partial P({\bf v},{\bf w})}{\partial t}&=&\frac{1}{2}
\int  d\hat{\bf n}\iiiint
d{\bf v}_ad{\bf w}_ad{\bf v}_bd{\bf w}_b\,
|({\bf v}_a-{\bf v}_b)\cdot \hat{\bf n}|^\gamma\,
P({\bf v}_a,{\bf w}_a)P({\bf v}_b,{\bf w}_b)\\
&\times&\big[\delta({\bf v}-{\bf v}_a')\delta({\bf w}-{\bf w}_a')
+\delta({\bf v}-{\bf v}_b')\delta({\bf w}-{\bf w}_b') -\delta({\bf
v}-{\bf v}_a)\delta({\bf w}-{\bf w}_a) -\delta({\bf v}-{\bf
v}_b)\delta({\bf w}-{\bf w}_b)\big].\nonumber
\end{eqnarray}
We integrate over all impact directions with $\int d\hat{\bf n}=1$
\cite{condition} and over the precollision velocities weighted by
the respective probability distributions. There are two gain terms
and two loss terms, because the velocities of interest $({\bf
v},{\bf w})$ can be identified with any one of the four velocities
in the collision rule \eqref{collision} and the kernel is simply the
collision rate \eqref{rate}.

\section{The Linear Kinetic Theory}

The focus of this study is the energy distribution that generally
depends only on two variables: $E_v$ and $E_w$. It is our aim to
compute the distribution $P(E_v,E_w)$ for asymptotically large
energies. This will be done for a system which is driven at very high
energies as well as for an undriven system.

As a first step to this goal, we simplify the Boltzmann equation in
the limit of large energies. Extremely energetic particles are rare
and as a result it is unlikely that such particles will encounter each
other.  Hence, energetic particles typically collide with much slower
particles. Since the collision rules are linear, the velocity of the
slower particle barely affects the outcome of the collision.  We can
therefore neglect the slower velocity.  Substituting \hbox{$({\bf
v}_a,{\bf w}_a)=({\bf v}_0,{\bf w}_0)$} and \hbox{$({\bf v}_b,{\bf
w}_b)=({\bf 0},{\bf 0})$} or \hbox{$({\bf v}_a,{\bf w}_a)=({\bf
0},{\bf 0})$} and \hbox{$({\bf v}_b,{\bf w}_b)=({\bf v}_0,{\bf w}_0)$}
into (\ref{rule}) gives the cascade process \cite{bm,bmm}
\begin{equation}
\label{cascade} ({\bf v}_0,{\bf w}_0)\to ({\bf v}_1,{\bf w}_1)+({\bf
v}_2,{\bf w}_2)
\end{equation}
where $({\bf v}_0,{\bf w}_0)$ is the precollision velocity of the
energetic particle and $({\bf v}_i,{\bf w}_i)$ with $i=1,2$ are the
consequent postcollision velocities. With these definitions, the
collision rules for extremely energetic particles are
\begin{subequations}
\label{rule-v}
\begin{align}
{\bf v}_1&=(1-\eta_n){\bf v}_0\cdot\hat{\bf n}\,\hat{\bf n}+
(1-\eta_t)({\bf v}_0-{\bf v}_0\cdot\hat{\bf n}\,\hat{\bf
n})-\eta_t\hat{\bf n}\times{\bf w}_0 \qquad
{\bf w}_1=\left(1-\frac{\eta_t}{q}\right){\bf w}_0+
\frac{\eta_t}{q}\hat{\bf n}\times{\bf v}_0\\
{\bf v}_2&=\eta_n{\bf v}_0\cdot\hat{\bf n}\,\hat{\bf n}+ \eta_t({\bf
v}_0-{\bf v}_0\cdot\hat{\bf n}\,\hat{\bf n})+\eta_t\hat{\bf
n}\times{\bf w}_0\quad\qquad\qquad\qquad \,{\bf
w}_2=-\frac{\eta_t}{q}{\bf w}_0+\frac{\eta_t}{q}\hat{\bf
n}\times{\bf v}_0,
\end{align}
\end{subequations}
where we have set $R=1$, so that the moment of inertia, $I=q$, is
dimensionless.  A collision between a high-energy particle and a
typical-energy particle produces two energetic particles with an
energy total that is smaller than the initial energy. This cascade
process transfers energy from large scales to small scales.

Since the cascade process \eqref{cascade} involves only one
particle, the tail of the probability distribution $P({\bf v},{\bf
w})$ obeys the {\it linear} equation
\begin{eqnarray}
\label{linear-v} \frac{\partial P({\bf v},{\bf w})}{\partial
t}\!=\!\! \iiint d\hat{\bf n} d{\bf v}_0d{\bf w}_0 |{\bf v}_0\cdot
\hat{\bf n}|^\gamma P({\bf v}_0,{\bf w}_0) \big[\delta({\bf
v}\!-\!{\bf v}_1)\delta({\bf w}\!-\!{\bf w}_1) \!+\!\delta({\bf
v}\!-\!{\bf v}_2)\delta({\bf w}\!-\!{\bf w}_2) \!-\!\delta({\bf
v}\!-\!{\bf v}_0)\delta({\bf w}\!-\!{\bf w}_0)\big].
\end{eqnarray}
There are two gain terms and one loss term according to the cascade
process \eqref{cascade}. Formally, this linear rate equation can be
obtained from the full nonlinear equation \eqref{nonlinear} by
treating either one of the precollision velocities as negligible and
then integrating over this small velocity.  This procedure leads to
four gain terms and two loss terms and thus, the factor $1/2$ in
\eqref{nonlinear} drops out. We stress that the linear equation
\eqref{linear-v} is valid only in the high-energy limit.

We also comment that the linear equation \eqref{linear-v} for the
high-energy tail of the velocity distribution may be valid in cases
where the full nonlinear equation is not. Whereas the nonlinear
equation requires that all possible velocities are uncorrelated, the
linear equation merely requires that energetic particles are
uncorrelated with typical particles. This is a much weaker condition.

In this paper, we restrict ourselves to two space dimensions, i.e.
rotating disks.  In that case the rotational velocities are always
perpendicular to the linear velocities.  Thus, we conveniently denote
the unit vector in the tangential direction by $\hat{\bf t}$ and the
unit vector coming out of the plane by $\hat{\bf z}$, such that
$\hat{\bf n}\cdot \hat{\bf t}=0$ and $\hat{\bf n}\times \hat{\bf
t}=\hat{\bf z}$. The precollision velocities of the energetic particle
${\bf v}_0=v_n\,\hat{\bf n}+v_t\,\hat{\bf t}$ and ${\bf
w}_0=w\,\hat{\bf z}$ are compactly written as $[v_n,v_t,w]$.  With
this notation, the postcollision velocities specified in
\eqref{linear-v} are
\begin{equation}
\Big[(1-\eta_n)v_n,(1-\eta_t)v_t+\eta_t w,(\eta_t/q)v_t+(1-\eta_t/q)w\Big],
\qquad{\rm and}\qquad
\Big[\eta_n v_n,\eta_t (v_t-w),(\eta_t/q)(v_t-w)\Big],
\end{equation}
respectively. We now treat the three velocity components, namely the
normal component of the velocity $v_n$, the tangential component of
the velocity $v_t$, and the scaled angular velocity $\sqrt{q}w$ as a
three dimensional vector with magnitude $V_0$, polar angle
$\theta_0$, and azimuthal angle $\phi_0$:
\begin{equation}
\label{vector}
(v_n,v_t,\sqrt{q}w)=
(V_0\sin\theta_0\cos\phi_0, V_0\sin\theta_0\sin\phi_0,V_0\cos\theta_0).
\end{equation}
The magnitude $V_0$ gives the energy
$E_0=\frac{1}{2}V_0^2=\frac{1}{2}(v_n^2+v_t^2+qw^2)$ while the polar
angle characterizes the fraction of energy stored in the rotational
degree of freedom, $\frac{1}{2}qw^2/E_0=\cos^2\theta$.  In this
representation, the postcollision velocities are three-dimensional
vectors with magnitude $V_i$, polar angle $0\leq \theta_i\leq \pi$,
and azimuthal angle $0\leq \phi_i\leq 2\pi$. The collision rules
\eqref{rule-v} allow us to express these quantities in terms of
$V_0,\theta_0,\phi_0$:
\begin{equation}
\label{abc-def}
(V_i\sin\theta_i\cos\phi_i,V_i\sin\theta_i\sin\phi_i,V_i\cos\theta_i)=
(V_0A_i,V_0B_i,V_0C_i)
\end{equation}
where $i=1,2$.  The magnitudes of the postcollision velocities are
proportional to the magnitude of the precollision velocity. The
three velocity components are scaled by three dimensionless constants
$A_i$, $B_i$ and $C_i$, that depend on the angles $\theta_0$ and
$\phi_0$ of the energetic particle, the collision parameters $\eta_n$
and $\eta_t$, and the moment of inertia $q$,
\begin{subequations}
\label{abc}
\begin{align}
A_1&=(1-\eta_n)\sin\theta_0\cos\phi_0\\
B_1&=(1-\eta_t)\sin\theta_0\sin\phi_0
+(\eta_t/\sqrt{q})\cos\theta_0\\
C_1&=(\eta_t/\sqrt{q})\sin\theta_0\sin\phi_0
+(1-\eta_t/q)\cos\theta_0
\\
A_2&=\eta_n\sin\theta_0\cos\phi_0\\
B_2&=\eta_t\sin\theta_0\sin\phi_0
-(\eta_t/\sqrt{q})\cos\theta_0\\
C_2&=(\eta_t/\sqrt{q})\sin\theta_0\sin\phi_0
-(\eta_t/q)\cos\theta_0.
\end{align}
\end{subequations}

The new energies are proportional to the precollision energies
\begin{equation}
\label{e-rules} E_i=\alpha_i E_0,\qquad {\rm with}\qquad
\alpha_i=A_i^2+B_i^2+C_i^2.
\end{equation}
We term the parameters $0<\alpha_i<1$ the contraction parameters.
Since the collisions are dissipative, these  parameters
satisfy the inequality $\alpha_1+\alpha_2\leq 1$. The equality
$\alpha_1+\alpha_2=1$ holds only for elastic collisions
($r_n=|r_t|=1$). The energy dissipation is $\Delta
E=E_0-E_1-E_2=\Lambda E$ with $\Lambda=1-\alpha_1-\alpha_2$ or
explicitly,
\begin{eqnarray}
\label{Lambda} \Lambda=\frac{1-r_n^2}{4}\sin^2\theta_0\cos^2\phi_0
+\frac{q}{1+q}\frac{1-r_t^2}{4}\Big(\sin^2\theta_0\sin^2\phi_0+
\frac{1}{q}\cos^2\theta_0\Big).
\end{eqnarray}
The polar and azimuthal angles are given by
\begin{eqnarray}
\label{omega-rules} \cos\theta_i=
\frac{C_i}{\sqrt{A_i^2+B_i^2+C_i^2}}\qquad {\rm and}
\qquad\tan\phi_i= \frac{B_i} {A_i},
\end{eqnarray}
respectively.

Let us represent solid angles by $\Omega\equiv \cos\theta,\phi$. With
this definition, the cascade process \eqref{rule-v} is
\begin{equation}
\label{cascade-e} (E_0,\Omega_0)\to (E_1,\Omega_1)+(E_2,\Omega_2)
\end{equation}
with $E_i$ and $\Omega_i$ given by \eqref{e-rules} and
\eqref{omega-rules}. Energetic particles have an important property:
the solid angle is not coupled to the energy!  Indeed, the
postcollision angles depend only on the precollision angle. The
cascade process has the following geometric interpretation: a three
dimensional vector is duplicated into two vectors. Subsequently,
these two vectors are scaled down by the contraction parameters
\eqref{e-rules}, and rotated according to the angular transformation
\eqref{omega-rules}.

We can now write the linear Boltzmann equation for $P(E,\Omega)$, the
distribution of energy and solid angle, in a {\it closed} form
\begin{eqnarray*}
\!\frac{\partial P(\!E,\!\Omega)}{\partial t}\!= \!\!\!\iint
\!\!dE_0 d\Omega_0 \big|\!\sqrt{E_0}\!\sin\theta_0\!\cos
\phi_0\!\big|^\gamma\!
P(\!E_0,\!\Omega_0)\!\big[\delta(E\!-\!E_1)\delta(\Omega\!-\!\Omega_1)\!+\!
\delta(E\!-\!E_2)\delta(\Omega\!-\!\Omega_2)\!
-\!\delta(E\!-\!E_0)\delta(\Omega\!-\!\Omega_0)\big].
\end{eqnarray*}
Time was rescaled, $t\to 2^{\gamma/2}t$, to absorb the constant which
arises from replacing velocity by energy in the collision rate
\eqref{rate}. Henceforth, we implicitly assume that the distribution
$P(E,\Omega)$ is independent of $\phi$ because the distribution of
{\it linear} velocities must be isotropic.  The integration over the
energy is performed using the collision rule \eqref{e-rules}, leading
to the linear rate equation for the tail of the energy distribution
\begin{equation}
\label{linear}
\frac{\partial P(E,\Omega)}{\partial t}\!=\! E^{\gamma/2}\!\!\int d\Omega_0
\big|\sin\theta_0\cos\phi_0\big|^\gamma\!
\left[\!P\left(\!\frac{E}{\alpha_1},\Omega_0\!\right)
\frac{\delta(\Omega-\Omega_1)}{\alpha_1^{1+\gamma/2}}\!+\!
      P\left(\!\frac{E}{\alpha_2},\Omega_0\!\right)
\frac{\delta(\Omega-\Omega_2)}{\alpha_2^{1+\gamma/2}}\!-\!
      P(E,\Omega_0)\delta(\Omega-\Omega_0)\right].
\end{equation}\
This is a non-local equation as the density of particles with energy
$E$ is coupled to the density of particles with the higher energies
$E/\alpha_1$ and $E/\alpha_2$. We stress that this equation is a
straightforward consequence of the cascade process \eqref{cascade-e}
and that it can also be derived from the full nonlinear Boltzmann
equation. Yet, there may be situations where the linear equation
\eqref{linear} is valid, while the nonlinear equation
\eqref{nonlinear} is not valid. The bivariate energy distributions
$P(E,\Omega)$ and $P(E_v,E_w)$ are completely equivalent but we
analyze the former because the cascade process \eqref{cascade-e} is
transparent in terms of the total energy and the solid angle.

\section{Driven Steady-states}

The inelastic Boltzmann equation admits stationary solutions for
{\it frictionless} particles. These stationary solutions describe
driven steady-states with rare but powerful injection of energy. The
injected energy cascades from high-energies down to small energies,
thereby balancing the energy lost in collisions. At energies below
the injection scale, Eqs.~\eqref{nonlinear}, \eqref{linear-v} and
\eqref{linear} are not altered by the energy source and
consequently, the stationary solution of the inelastic Boltzmann
equation holds up to this large energy scale \cite{bm,bmm}.  Here,
we seek a corresponding stationary solution for particles with {\it
rotational} degrees of freedom in the high energy limit.

The stationary solution has to fulfill Eq.~\eqref{linear} with the left
hand side set to zero
\begin{equation}
\label{stationary-eq}
0=
\int d\Omega_0
\big|\sin\theta_0\cos\phi_0\big|^\gamma\!
\left[\!\frac{1}{\alpha_1^{1+\gamma/2}}\!P\left(\!\frac{E}{\alpha_1},
\Omega_0\!\right)
\delta(\Omega-\Omega_1)\!+\!
\!\frac{1}{\alpha_2^{1+\gamma/2}}\!P\left(\!\frac{E}{\alpha_2},
\Omega_0\!\right)
\delta(\Omega-\Omega_2)\!-\!
      P(E,\Omega_0)\delta(\Omega-\Omega_0)\right].
\end{equation}
At high-energies, the solid angle is not coupled to the energy, as
follows from Eq.~\eqref{omega-rules}. This fact has a major
consequence: the bivariate energy distribution $P(E,\Omega)$ takes the
form of a product of the energy distribution \hbox{$p(E)=\int
d\Omega\, P(E,\Omega)$} and the distribution of solid angle,
$g(\Omega)$,
\begin{equation}
\label{product} P(E,\Omega)\to p(E)\,g(\Omega)
\end{equation}
as $E\to\infty$. The angle distribution is normalized, $\int
d\Omega\, g(\Omega)=1$. It does not depend on the
azimuthal angle, because on average the two components of the linear
velocity are equivalent. Due to the equi-dimensional (in $E$)
structure of the steady-state equation \eqref{stationary-eq}, the
product ansatz \eqref{product} is a solution when the distribution
$p(E)$ decays algebraically
\begin{equation}
\label{powerlaw}
p(E)\sim E^{-\nu},
\end{equation}
as $E\to\infty$ \cite{bm,bmm}.  We obtain a closed equation for the
distribution $g(\Omega)$ by substituting the product ansatz
\eqref{product} with the power-law form \eqref{powerlaw} into the
steady-state equation \eqref{stationary-eq}
\begin{equation}
\label{g-eq}
0=\int d\Omega_0\, g(\Omega_0)\,
\big|\sin\theta_0\cos\phi_0\big|^\gamma
\left[\alpha_1^{\nu-1-\gamma/2}\delta(\Omega-\Omega_1)+
      \alpha_2^{\nu-1-\gamma/2}\delta(\Omega-\Omega_2)-
\delta(\Omega-\Omega_0)\right].
\end{equation}
This equation is linear in $g(\Omega)$.  However, it is nonlinear in
$\nu$ and moreover, the solid angles
$\Omega_{i}\equiv\Omega_{i}(\Omega_0)$ in \eqref{omega-rules} and the
contraction parameters $\alpha_i\equiv\alpha_i(\Omega_0)$ in
\eqref{e-rules} are complicated functions of the solid angle
$\Omega_0$.

Equation (\ref{g-eq}) involves two unknowns quantities, the exponent
$\nu$ and the distribution function $g(\Omega)$. A solution does not
exist for arbitrary values of $\nu$. In fact, there is one and only
one value of $\nu$ for which there is a solution for $g(\Omega)$. This
is the value selected by the cascade dynamics!  In other words,
\eqref{g-eq} is an eigenvalue equation: $\nu$ is the eigenvalue and
$g(\Omega)$ is the eigenfunction. This eigenvalue equation circumvents
the full nonlinear equation \eqref{nonlinear} and thus, represents a
significant simplification.

The physical interpretation of \eqref{g-eq} involves a cascade
process in which the solid angle undergoes a creation-annihilation
process
\begin{equation}
\label{angular}
\Omega_0\to
\begin{cases}
\emptyset& {\rm with\ rate\ } \beta_0,\\
\Omega_1& {\rm with\ rate\ } \beta_1,\\
\Omega_2& {\rm with\ rate\ } \beta_2.
\end{cases}
\end{equation}
Here, $\beta_i=|\sin\theta_0\cos\phi_0\big|^\gamma\alpha_i$ for
$i=0,1,2$ and $\alpha_0=1$. There is one annihilation process and two
creation processes. These processes have relative weights that reflect
the powerlaw decay of $p(E)$. At the steady-state, the creation and
the annihilation terms balance (see Appendix \ref{stationary}), as
reflected in the integrated form of \eqref{g-eq}
\begin{equation}
\label{balance} 0=\int d\Omega_0\, g(\Omega_0)
\big|\sin\theta_0\cos\phi_0\big|^\gamma
\left[\alpha_1^{\nu-1-\gamma/2}+
      \alpha_2^{\nu-1-\gamma/2}-1\right].
\end{equation}
To achieve a steady-state, $\beta_i<\beta_0$ for $i=1,2$ and therefore
$\alpha_i^{\nu-1-\gamma/2}<1$. Since $\alpha_i<1$, we have the lower
bound $\nu>1+\gamma/2$.

\begin{figure}[t]
\includegraphics*[width=0.5\textwidth]{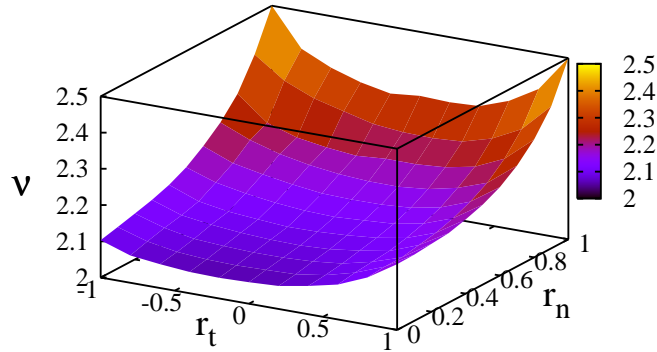}
\caption{The exponent $\nu$ for hard spheres ($\gamma=1$) as a
  function of the coefficients of normal, $r_n$, and tangential,
  $r_t$, restitution coefficients. The numerical procedure for solving
  \eqref{g-eq} is detailed below.}
\label{exponent}
\end{figure}

We can immediately check that for elastic collisions,
$\nu=2+\gamma/2$ \cite{avb,asb} because $\alpha_1+\alpha_2=1$, and
therefore, we conclude the bounds $1+\gamma/2\leq\nu\leq
2+\gamma/2$.  The exponent $\nu$ varies continuously with the
restitution coefficients $r_n$ and $r_t$ and the normalized moment
of inertia $q$.  This quantity must coincide with the value found
for frictionless particles where tangential restitution is
irrelevant ($r_t=-1$) \cite{bm,bmm,relate}, but otherwise the
exponent is distinct, as shown in Fig. \ref{exponent}. Also, the
exponent $\nu$ increases monotonically with $r_n$ and $|r_t|$. We
conclude that the rotational degrees of motion do affect the
power-law behavior \eqref{powerlaw}.

The azimuthal angle $\theta$ characterizes the fraction of energy
stored in the rotational mode, $\cos^2 \theta=E_w/E$ with
$E_w=\frac{1}{2}qw^2$. The angle distribution
$g(\Omega)=(2\pi)^{-1}\tilde f(\cos\theta)$ therefore captures the
partition of energy into rotational and translational energies. We
introduce the natural variable $0\leq x\leq 1$ defined by
$x=|\cos\theta|$ so that
\begin{equation}
\label{x-def} x=\sqrt{\frac{E_w}{E}}
\end{equation}
and present results for the angle distribution $f(x)=2\tilde
f(\cos\theta)$.  In equilibrium, energy is partitioned equally into
all degrees of freedom and therefore $g_{\rm eq}(\Omega)=(4\pi)^{-1}$
or equivalently,
\begin{equation}
\label{feq}
f_{\rm eq}(x)=1
\end{equation}
for $0\leq x\leq 1$. In particular, $\langle x^2\rangle=1/3$.

\subsection{Simulation Methods}

We numerically studied the angle distribution $f(x)$ by solving the
linear eigenvalue equation \eqref{g-eq} for the ``angular'' process
\eqref{angular} and by solving the full nonlinear Boltzmann equation
\eqref{nonlinear} for the collision process \eqref{collision}. Both of
these equations are solved using Monte Carlo simulations.

The eigenvalue equation is solved by mimicking the angular process.
Throughout the simulation, the value $\nu$ is fixed. There are $N$
particles, each with a given polar angle. A particle with polar angle
$\theta_0$ is picked at random and then, a random azimuthal angle
$\phi_0$ is drawn. The polar angles $\theta_1$ and $\theta_2$ are then
calculated according to \eqref{omega-rules}. The original particle is
annihilated with probability $\beta_0$ and simultaneously, a new
particle with angle $\theta_1$ is created with probability $\beta_1$
and similarly, a second particle with angle $\theta_2$ is created with
probability $\beta_2$. Therefore, the number of particles may increase
by one, remain unchanged, or decrease by one. The exponent $\nu$ is
the value that keeps the total number of particles constant in the
long time limit. The eigenvalue $\nu$ is calculated by repeating this
simulation for various values of $\nu$ and then using the bisection
method \cite{ptvf}. We present Monte Carlo simulations of $100$
independent realizations with $N=10^7$ particles.

Driven steady-states are obtained by simulating the two competing
processes of inelastic collisions and energy injection. In an
inelastic collision, two particles are picked at random and also,
the impact direction is chosen at random. The particle velocities
are updated according to the collision law \eqref{rule}.  Collisions
are executed with probability proportional to the collision rate.
Throughout this process, we keep track of the total energy loss.
With a small rate, we augment the energy of a randomly selected
particle by an amount equal to the loss total and subsequently,
reset the total energy loss to zero. A fraction of the injected
energy is rotational and the complementary fraction is
translational. We draw this fraction according to the equilibrium
distribution \eqref{feq}. We experimented with different angle
distributions and found that the resulting stationary state did not
change.

Obtaining the distribution $f(x)$ is generally challenging as it
requires excellent statistics. The simulations are most efficient for
Maxwell molecules because all possible collisions are equally likely.
Therefore, for the full nonlinear Boltzmann equation
\eqref{nonlinear}, we present the angle distribution of the energetic
particles only for the case $\gamma=0$.

For Maxwell molecules, the injection rate is $10^{-4}$ and the system
size is $N=10^7$. The corresponding values for hard spheres are
$10^{-2}$ and $N=10^5$. In all cases, the simulation results represent
an average over $10^2$ independent realizations. Unless noted
otherwise, the simulation results are for maximally dissipative
($r_n=r_t=0$) disks ($q=1/2$).

\subsection{The Distribution of Total Energy}

\begin{figure}[t]
\begin{minipage}{0.47\textwidth}
\includegraphics*[width=\textwidth]{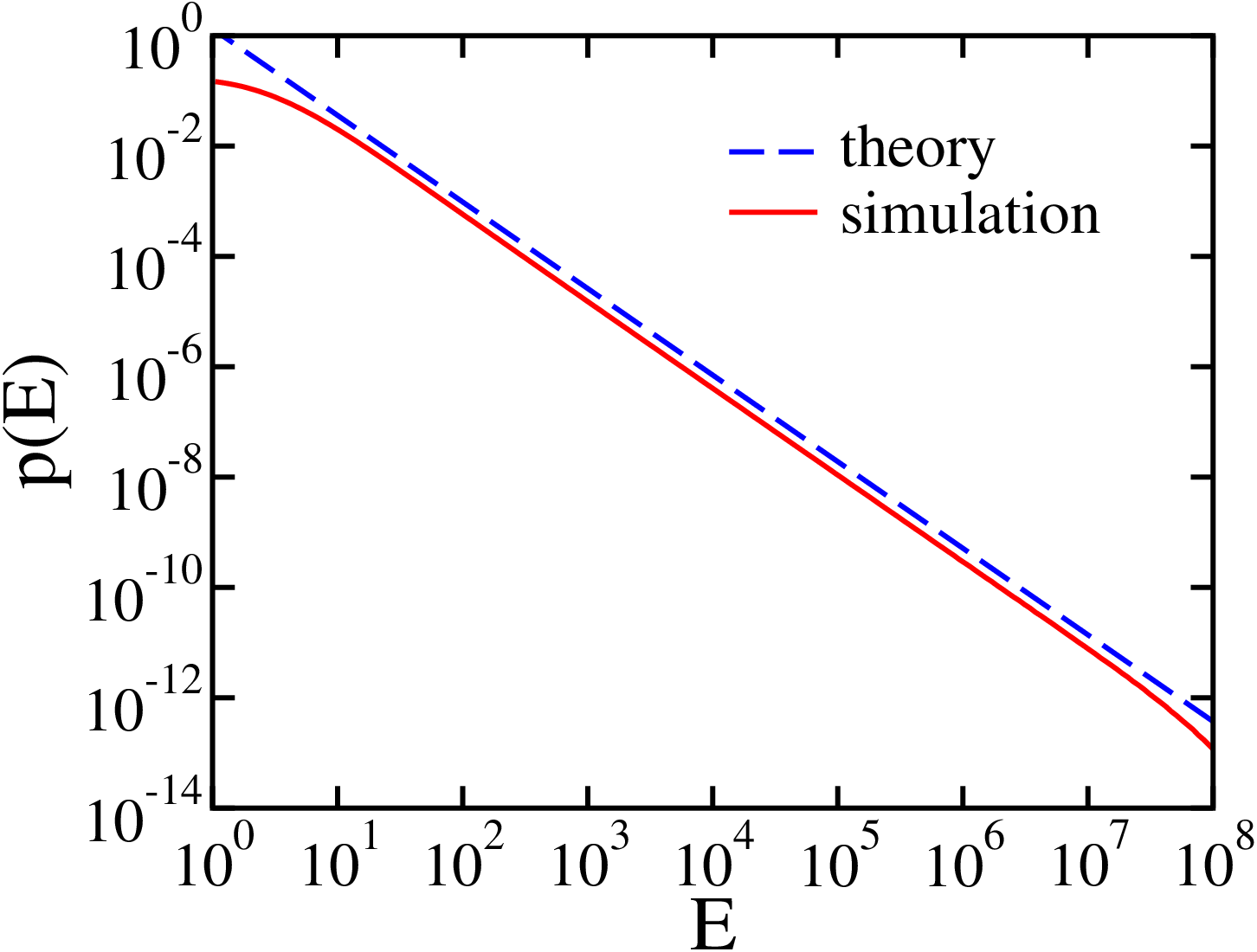}
\caption{The tail of the energy distribution for driven Maxwell
molecules. Shown are simulation results (solid line) and a line with
the slope predicted by the theory (dashed line). The energy is
normalized by the typical energy $10^{-4}$.\label{tail-Maxwell}}
\end{minipage}
\hfill
\begin{minipage}{0.47\textwidth}
\includegraphics*[width=\textwidth]{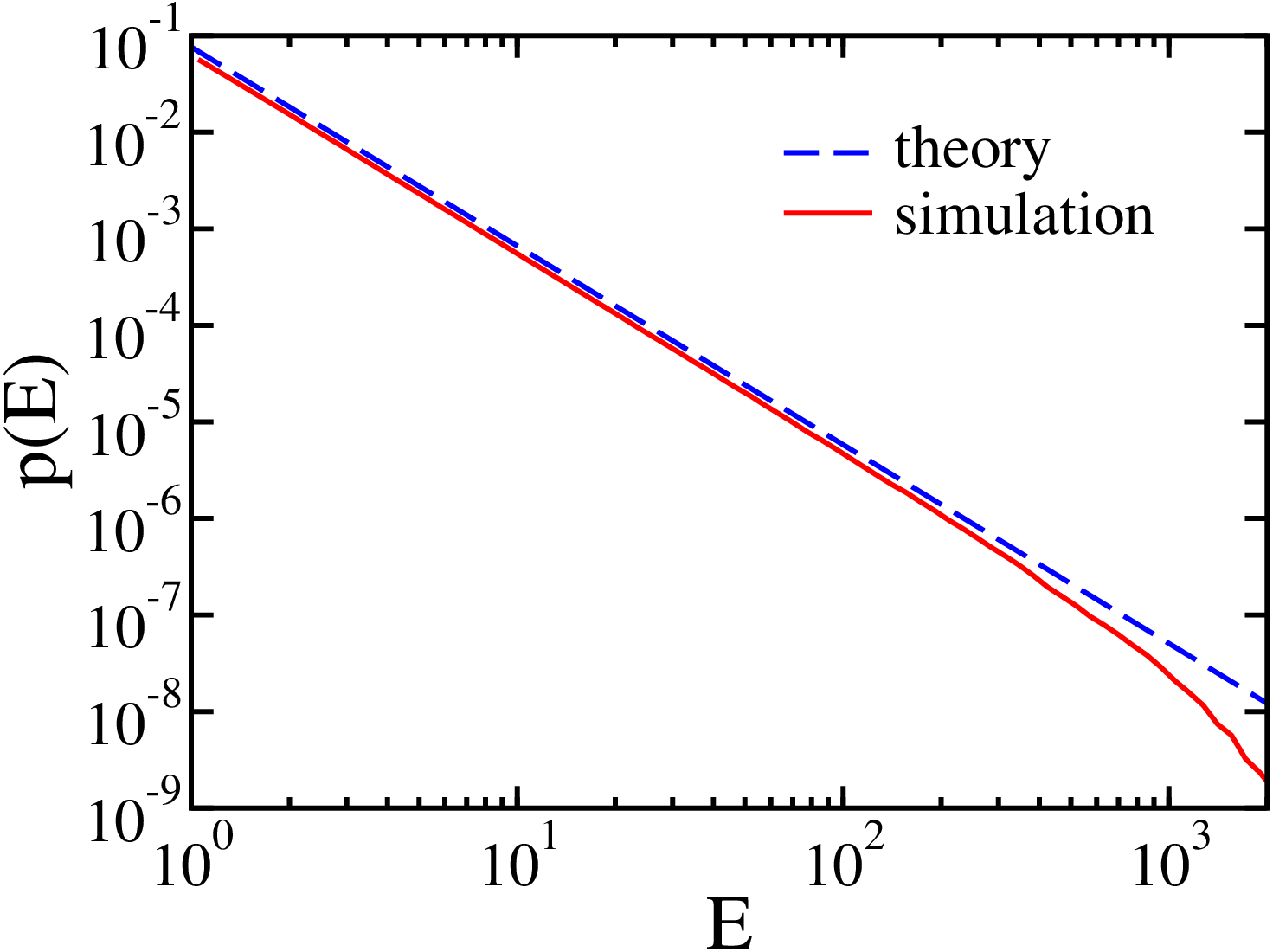}
\caption{The tail of
the energy distribution for driven hard spheres. Shown are
simulation results (solid line) and a line with the slope predicted
by the theory (dashed line).\label{tail-hard}}
\end{minipage}
\end{figure}

The numerical simulations confirm several of our theoretical
predictions. First, the energy distribution approaches a
steady-state with a power-law high-energy tail. Second, the
distribution of the total energy $p(E)$ decays algebraically as in
\eqref{powerlaw}. Third, the exponent $\nu$ is in excellent
agreement with the predictions of the eigenvalue equation. For
Maxwell molecules, Monte Carlo simulation of the full nonlinear
equation yields $\nu=1.570\pm0.005$ whereas numerical solution of
the eigenvalue equation \eqref{g-eq} gives $\nu=1.569\pm 0.005$
(Fig. \ref{tail-Maxwell}). For hard-spheres, where the simulation
results are slightly less accurate, the corresponding values are
$\nu=2.065\pm 0.005$ and $\nu=2.060\pm 0.005$ (Fig. \ref{tail-hard}). 
The behavior of the distribution of {\it total}
energy is therefore qualitatively similar to the behavior in the
no-rotation case \cite{bm,bmm}. However, the quantitative behavior
is different because the exponent $\nu$ does depend on the
tangential restitution coefficient and the moment of inertia 
(Fig. \ref{exponent}).

\subsection{The Angle distribution}

\begin{figure}[t]
\begin{minipage}{0.47\textwidth}
\includegraphics*[width=\textwidth]{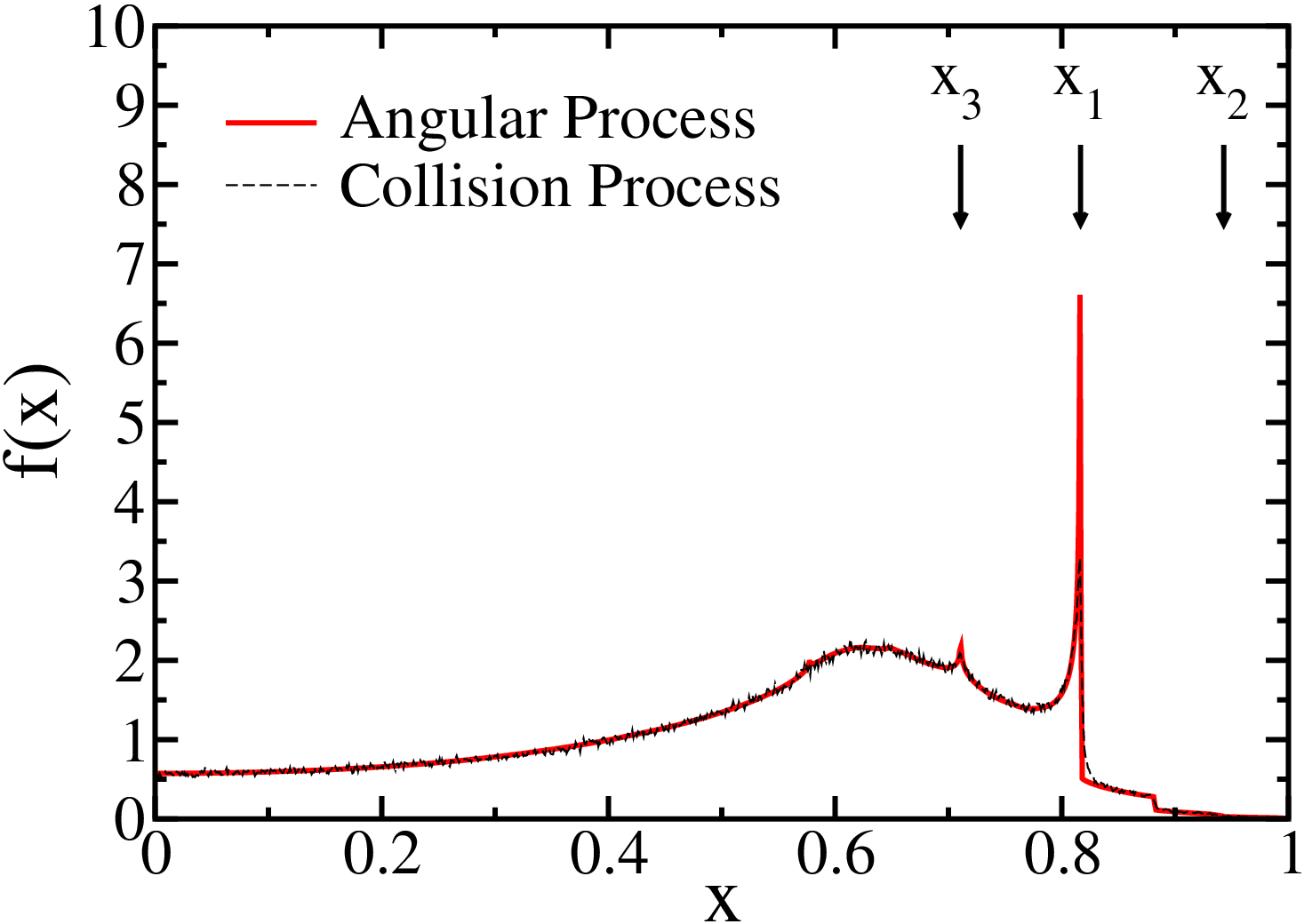}
\caption{The angle distribution $f(x)$ obtained by Monte Carlo
simulation of the angular process \eqref{angular} (solid line) and the
collision process \eqref{collision} (dashed line) for Maxwell
molecules. The special values $x_1$, $x_2$, and $x_3$ discusses in the
text are indicated by arrows.
\label{angle-Maxwell}}
\end{minipage}
\hfill
\begin{minipage}{0.47\textwidth}
\includegraphics*[width=\textwidth]{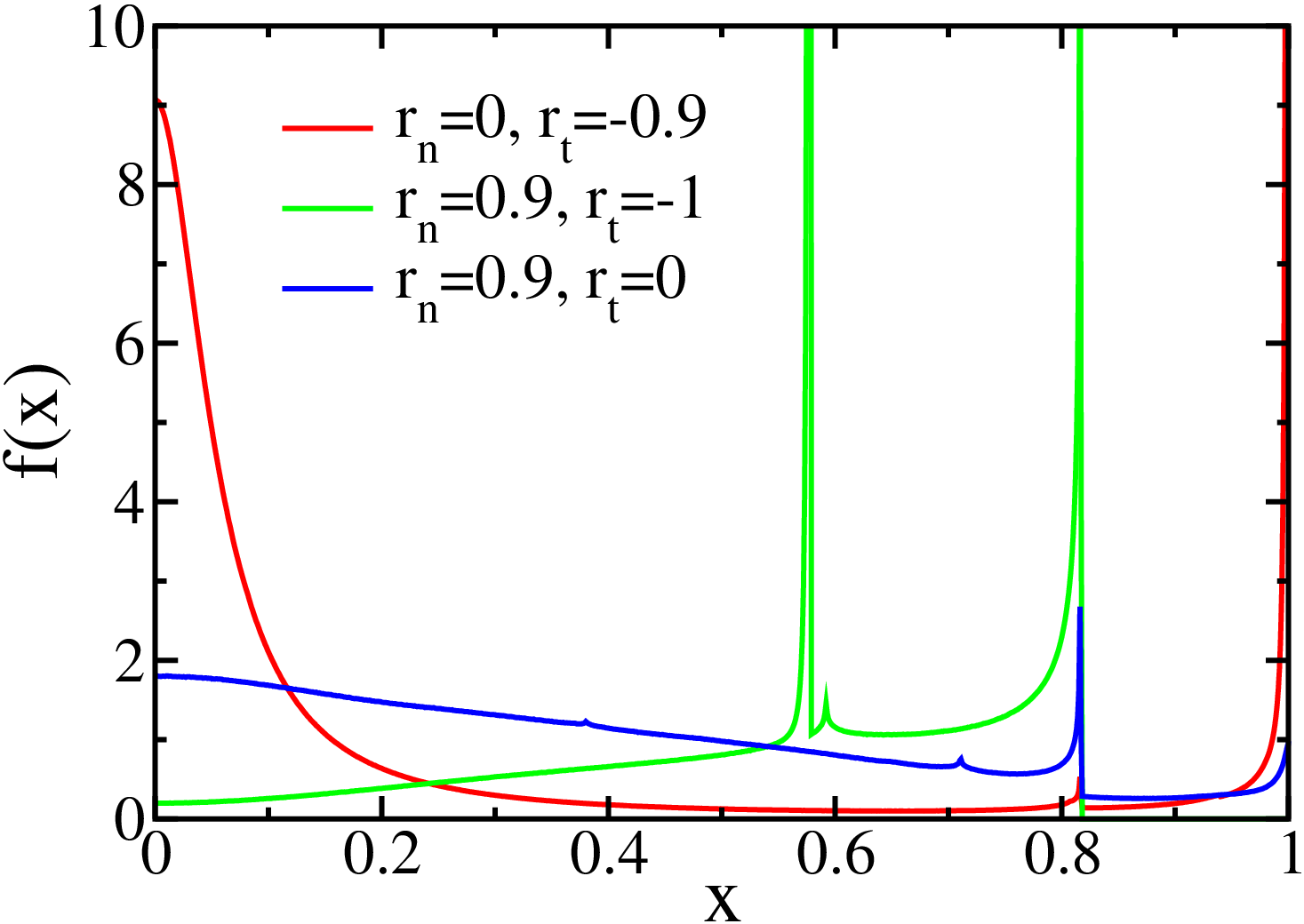}
\caption{The angle distribution $f(x)$ for various collision
parameters ($r_n$ and $r_t$) for Maxwell molecules.\label{angle-galery}}
\end{minipage}
\end{figure}

The numerical simulations also confirm several of our theoretical
predictions concerning the angle distribution.  Extremely energetic
particles have a {\it universal} distribution $f(x)$.  This
distribution is independent of the energy, provided that the energy is
sufficiently large. We had to probe only the most energetic particle
out of roughly $10^3$ particles to measure this distribution.  For
this reason, the linear analysis and the resulting eigenvalue equation
are valuable because they allow for an accurate and efficient
determination of the angle distribution of the energetic particles.
We also verified that the distribution $f(x)$ obeys the eigenvalue
equation \eqref{g-eq}, as demonstrated in Fig.~\ref{angle-Maxwell},
where the simulations are compared to the solution of the angular
process.

The distribution $f(x)$ has several noteworthy features. First, it
is not uniform, implying the breakdown of energy equipartition in a
granular gas. Furthermore, this distribution is {\it nonanalytic}.
It contains singularities and discontinuous derivatives. There are
notable peaks in the distribution so that special values $x$ and
special ratios $E_w/E$ are strongly preferred. The reason for these
peaks is the fact that the polar angle is limited. For example,
$\cos^2\theta_2<1/(1+q)$ as seen by substituting $\cos\theta_0=\pm1$
into \eqref{abc} and \eqref{omega-rules}. Consequently, there is a
special ratio
\begin{equation}
\label{special1} x_1=\sqrt{\frac{1}{1+q}}
\end{equation}
with the corresponding special energy ratio $E_w/E=x_1^2$.  This is
the most pronounced peak in Fig.~\ref{angle-Maxwell},
\hbox{$x_1=\sqrt{2/3}=0.81649$}. Numerically, we observe that the
peak becomes more pronounced as the distribution is measured at a
finer scale, indicating that the distribution function diverges at
this point.

Similarly, there is another special ratio that corresponds to
$\theta_1$ when $\cos\theta_0=\pm1$, and unlike \eqref{special1}, this
location depends on the tangential restitution,
\begin{equation}
\label{special2}
x_2=\frac{1-\eta_t/q}{\sqrt{\eta_t^2/q+(1-\eta_t/q)^2}}.
\end{equation}
Indeed, there is a barely noticeable cusp at
$x_2=\sqrt{8/9}=0.942809$. Singularities may induce less pronounced
``echo''-singularities.  For example, using $\cos\theta_0=x_1$ and
$\phi_0=\pi/2$ yields the special ratio
\begin{equation}
\label{special3} x_3=\frac {1+\eta_t(1-1/q)}
{\sqrt{q[1-\eta_t(1-1/q)]^2+[1+\eta_t(1-1/q)]^2}}.
\end{equation}
There is a noticeable peak at the corresponding value
$x_3=\sqrt{50/99}=0.710669$ in Fig.~\ref{angle-Maxwell}. We anticipate
that as the transformation \eqref{omega-rules} is iterated, the
strength of the singularities weakens and as a result there are
discontinuous derivatives of increasing order, a subtle behavior that
is difficult to measure.

\begin{figure}[t]
\begin{minipage}{0.47\textwidth}
\includegraphics*[width=\textwidth]{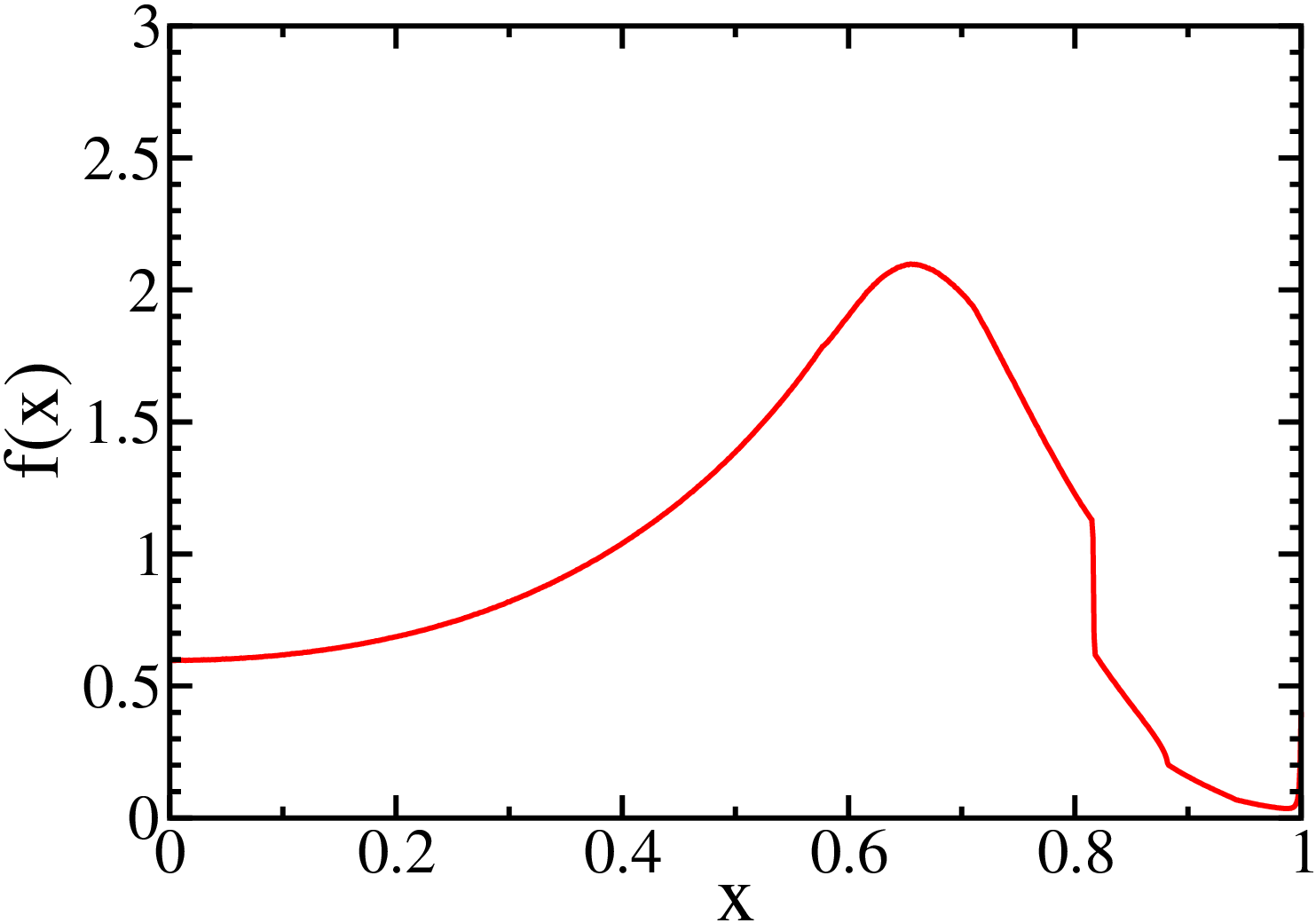}
\caption{The angle distribution $f(x)$ for hard spheres.\label{angle-hard}}
\end{minipage}
\hfill
\begin{minipage}{0.47\textwidth}
\includegraphics*[width=\textwidth]{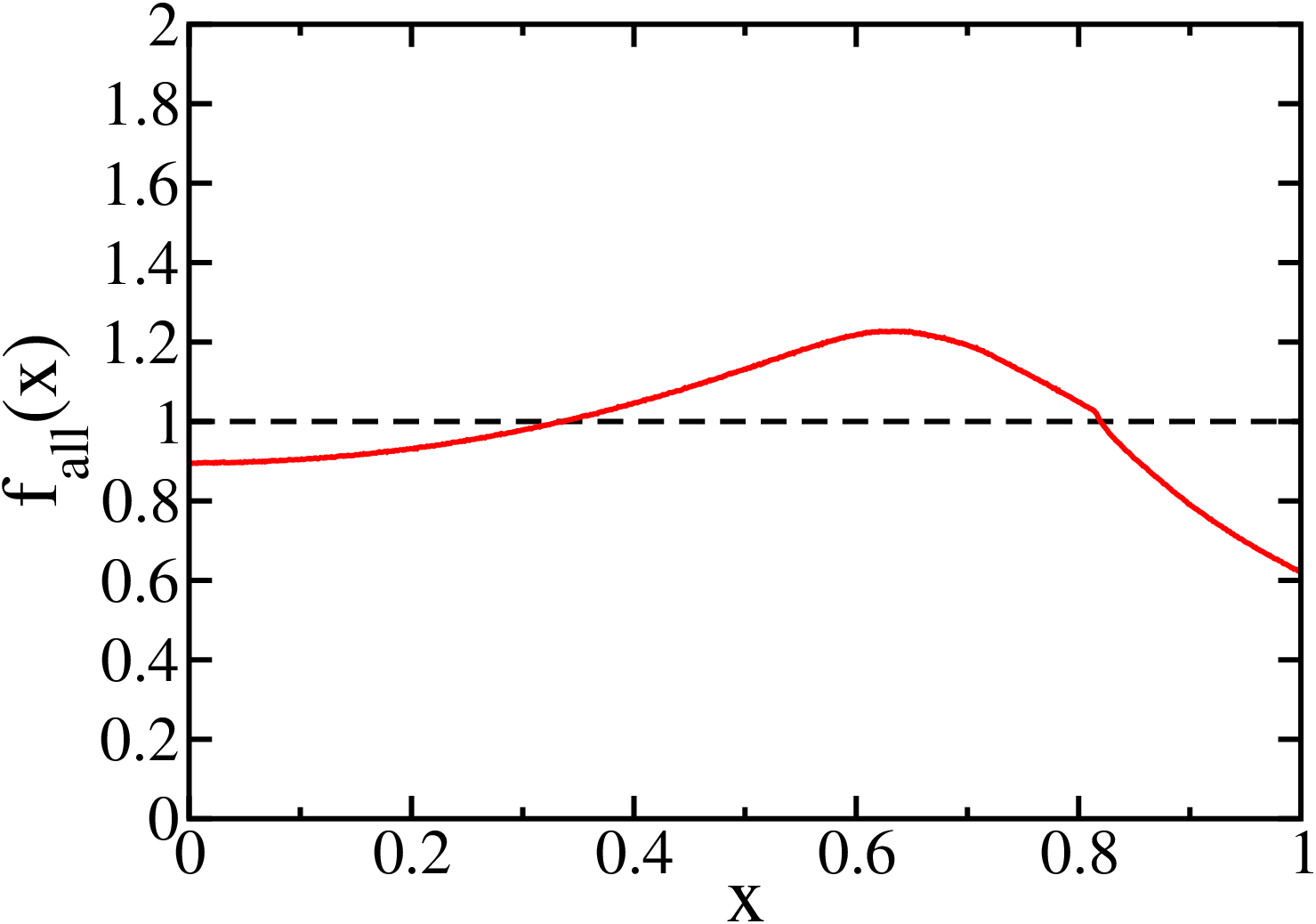}
\caption{The angle distribution $f_{\rm all}(x)$ of all particles
for Maxwell molecules (solid line). Also shown for reference is the
uniform equilibrium distribution (broken line).\label{angle-all}}
\end{minipage}
\end{figure}

The location of the singularities varies with the collision parameters
$r_n$ and $r_t$ and the moment of inertia $q$. In fact, the angle
distribution is extremely sensitive to material properties as its
shape changes dramatically with these parameters, see
Fig.~\ref{angle-galery}. The angle distribution also depends on the
collision rate and it is much smoother for hard spheres, see
Fig.~\ref{angle-hard}.  Since the collision rate vanishes for grazing
collisions, $\phi=\pi/2$, the associated singularities including in
particular \eqref{special3} are suppressed.  Nevertheless, there is a
pronounced jump at the special ratio given by \eqref{special1} and
there are also noticeable cusps.

The angle distribution of all particles $f_{\rm all}(x)\propto\int
dE\,P(E,\Omega)$ is shown in Fig.~\ref{angle-all}. It is
substantially different from $f(x)$. Therefore, the energy
distribution $P(E,\Omega)$ does not factorize in general and there
are correlations between the solid angle and the total energy. Only
for energetic particles does \eqref{product} hold. Moreover, $f_{\rm
all}(x)$ is much smoother in comparison with $f(x)$ although there
is a jump in the first derivative at the special ratio
\eqref{special1} showing that the angle distribution of all
particles is also non-analytic, see Fig.~\ref{angle-all}. Generally,
the angle distribution depends on energy and the deviation from a
uniform distribution grows with energy.

We also comment that lone measurement of the moment $\langle
x^2\rangle$ can be misleading. The angle distribution may very well
have a value close to the equipartition value $\langle x^2\rangle_{\rm
eq}=1/3$ but still, be very far from the equilibrium
distribution. Indeed, in Fig. \ref{angle-Maxwell}, $\langle
x^2\rangle\cong 0.318$, a value that barely differs from the
equilibrium value, even though the corresponding distribution is far
from uniform. The second moment may also differ substantially from the
equipartition value and for example, $\langle x^2\rangle=0.202$ when
$r_n=0.9$ and $r_t=0$ (Fig. \ref{angle-galery}).

We argue that the qualitative features of the angle distribution
should be generic in granular materials. Collisions involving
energetic particles must follow the linear cascade rules
\eqref{cascade-e} with the angular transformations
\eqref{omega-rules}. The singularities are a direct consequence of
these transformations and therefore should be generic. Measuring the
parameter-sensitive distribution $f(x)$ experimentally is
challenging because a huge number of particles must be probed and
the measurement has to be accurate. The distribution $f_{\rm
all}(x)$ provides a detailed probe of the partition of energy into
rotational and translation motion.

\section{Free Cooling}

We now consider freely cooling granular gases that evolve via purely
collisional dynamics. Without energy input, all energy is eventually
dissipated and the particles come to rest. This system has been
studied extensively \cite{bp} for hard spheres with \cite{hz,bpkz}
and without rotation \cite{ep}.

We consider Maxwell molecules where in the absence of rotation an
exact treatment is possible \cite{kb,eb,bk,bcg,bc}. When
$\gamma=0$ the Boltzmann equation (\ref{nonlinear}) simplifies
\begin{eqnarray}
\label{nonlinear-mm} \frac{\partial P({\bf v},{\bf w})}{\partial
t}&=&\frac{1}{2} \int  d\hat{\bf n}\iiiint d{\bf v}_ad{\bf w}_ad{\bf
v}_bd{\bf w}_b\,
P({\bf v}_a,{\bf w}_a)P({\bf v}_b,{\bf w}_b)\\
&\times&[\delta({\bf v}-{\bf v}_a')\delta({\bf w}-{\bf w}_a')
+\delta({\bf v}-{\bf v}_b')\delta({\bf w}-{\bf w}_b')
-\delta({\bf v}-{\bf v}_a)\delta({\bf w}-{\bf w}_a)
-\delta({\bf v}-{\bf v}_b)\delta({\bf w}-{\bf w}_b)].\nonumber
\end{eqnarray}
Consequently, the equations for the moments $\langle v^n w^m
\rangle=\iint d{\bf v} d{\bf w}P({\bf v},{\bf w}) v^n w^m$ close.

\subsection{The Temperatures}

Here, we consider only the translational temperature defined as the
average translational energy, $T_v=\langle E_v\rangle$, and the
rotational temperature, defined as the average rotational energy
$T_w=\langle E_w\rangle$. These two temperatures are coupled through
the linear equation
\begin{equation}
\label{matrix}
\frac{d}{dt}
\left(\begin{array}{c}T_v \\ T_w \\\end{array}\right)=-
\left(%
\begin{array}{cc}
  \lambda_{vv} & \lambda_{vw} \\
  \lambda_{wv} & \lambda_{ww} \\
\end{array}%
\right)
\left(\begin{array}{c} T_v \\ T_w \\\end{array}\right).
\end{equation}
Appendix \ref{coefficents} details the derivation of the matrix of
coefficients
\begin{subequations}
\label{lambda}
\begin{align}
\lambda_{vv}&=\eta_n(1-\eta_n)+\eta_t(1-\eta_t)\\
\lambda_{vw}&=-2\eta_t^2/q\\
\lambda_{wv}&=-\eta_t^2/q\\
\lambda_{ww}&=2(\eta_t/q)(1-\eta_t/q).
\end{align}
\end{subequations}
The two temperatures are coupled as long as $\eta_t\neq 0$ or
alternatively, $r_t\neq -1$.

The solution of \eqref{matrix} is a linear combination of the two
eigenvectors
\begin{equation}
\left(\begin{array}{c}T_v \\ T_w \\\end{array}\right)=
C_-\left(
\begin{array}{c}
  1  \\
  c_- \\
\end{array}
\right)e^{-\lambda_-t}+
C_+\left(
\begin{array}{c}
  1 \\
  c_+ \\
\end{array}
\right)e^{-\lambda_+t}
\end{equation}
with the constants $C_-$ and $C_+$ set by the initial conditions,
and $c_{\pm}=(\lambda_\pm-\lambda_{vv})/\lambda_{vw}$. The
eigenvalues are
\begin{equation}
\label{eigenvalues}
\lambda_{\pm}=\frac{\lambda_{vv}+\lambda_{ww}}{2}\pm
\sqrt{\left(\frac{\lambda_{vv}-\lambda_{ww}}{2}\right)^2+
\lambda_{vw}{\lambda_{wv}}}\,.
\end{equation}
The larger eigenvalue is irrelevant in the long time limit and
therefore,
\begin{equation}
\left(\begin{array}{c}T_v \\ T_w \\\end{array}\right)\to
C_-\left(
\begin{array}{c}
  1  \\
  c_- \\
\end{array}
\right)e^{-\lambda t}
\end{equation}
such that both temperatures decay with the same rate
$\lambda\equiv\lambda_-$. Of course, the total temperature also
follows the same exponential decay, \hbox{$T=T_v+T_w\sim e^{-\lambda
t}$}. In this regime, the fraction of rotational energy is on
average
\begin{equation}
\lim_{t\to\infty}\frac{T_w}{T}=\frac{c_-}{1+c_-}=
\frac{\lambda-\lambda_{vv}}{\lambda+\lambda_{vw}-\lambda_{vv}}.
\end{equation}
The approach toward this value is exponentially fast and the
relaxation time is inversely proportional to the difference in
eigenvalues $\tau=1/(\lambda_+-\lambda_-)$.

In equilibrium, $T_w/T=1/3$ but for nonequilibrium granular gases the
ratio varies. In Fig.~8 we plot the ratio of the average rotational
energy to the total energy as a function of the coefficients of
restitution. In accordance with our findings for driven steady-states,
energy is not partitioned equally between all the degrees of freedom.

\subsection{The Energy Distribution}

To study the full energy distribution, it is again convenient to
make a transformation of variables from the velocity pair $({\bf
v},{\bf w})$ to the total energy and the solid angle $(E,\Omega)$.
The energy distribution is now time dependent and assuming that the
temperature -- $T\sim e^{-\lambda t}$ -- is the characteristic
energy scale we postulate the self-similar form
\begin{equation}
\label{scaling}
P(E,\Omega,t)\to e^{\lambda t}\Phi(Ee^{\lambda t},\Omega)
\end{equation}
with the prefactor ensuring proper normalization, $\iint dz\,d\Omega\,
\Phi(z,\Omega)=1$. We focus on the high-energy behavior where the
linear equation (\ref{linear}) holds. By substituting the scaling form
(\ref{scaling}) into this linear equation and setting $\gamma=0$, we
find the integro-differential equation governing the scaling function
\begin{eqnarray}
\label{Phi-eq}
\lambda\Phi(z,\Omega)+\lambda z\,\frac{d}{dz}\Phi(z,\Omega)\!=\!
\int\! d\Omega_0\!\left[
\frac{1}{\alpha_1}\Phi\left(\frac{z}{\alpha_1},\Omega_0\right)
\delta(\Omega-\Omega_1)+
\frac{1}{\alpha_2}\Phi\left(\frac{z}{\alpha_2},\Omega_0\right)
\delta(\Omega-\Omega_2)-\Phi(z,\Omega_0)\delta(\Omega-\Omega_0)\right].
\end{eqnarray}
We again write the multivariate energy distribution as a product
$\Phi(z,\Omega)\to \psi(z)g(\Omega)$ of the distribution of the total
energy $\psi(z)=\int d\Omega\, \Phi(z,\Omega)$ and the distribution of
the solid angle $g(\Omega)$. This form is a solution of the
equi-dimensional equation (\ref{Phi-eq}) when the distribution of the
total energy decays as a power-law
\begin{equation}
\psi(z)\sim z^{-\nu}
\end{equation}
at large energies, $z\to\infty$. The angle distribution satisfies
the eigenvalue equation
\begin{eqnarray}
\label{g-eq-1} 0=\int d\Omega_0 \,g(\Omega_0)
\Big\{\alpha_1^{\nu-1}\delta(\Omega-\Omega_1)+
\alpha_2^{\nu-1}\delta(\Omega-\Omega_2)-[1-\lambda(\nu-1)]
\delta(\Omega-\Omega_0)\Big\}.
\end{eqnarray}
Of course, setting $\lambda=0$, one recovers the steady-state equation
\eqref{g-eq} reflecting that the similarity solution is stationary.
The factor $1$ is replaced by the smaller factor $1-\lambda(\nu-1)$
that accounts for the constant decrease in the number of particles at
any given energy because of dissipation. Again, we have a nonlinear
eigenvalue equation with the eigenvalue $\nu$ and the eigenfunction
$g(\Omega)$.

\begin{figure}[t]
\begin{minipage}{0.49\textwidth}
\includegraphics*[width=\textwidth]{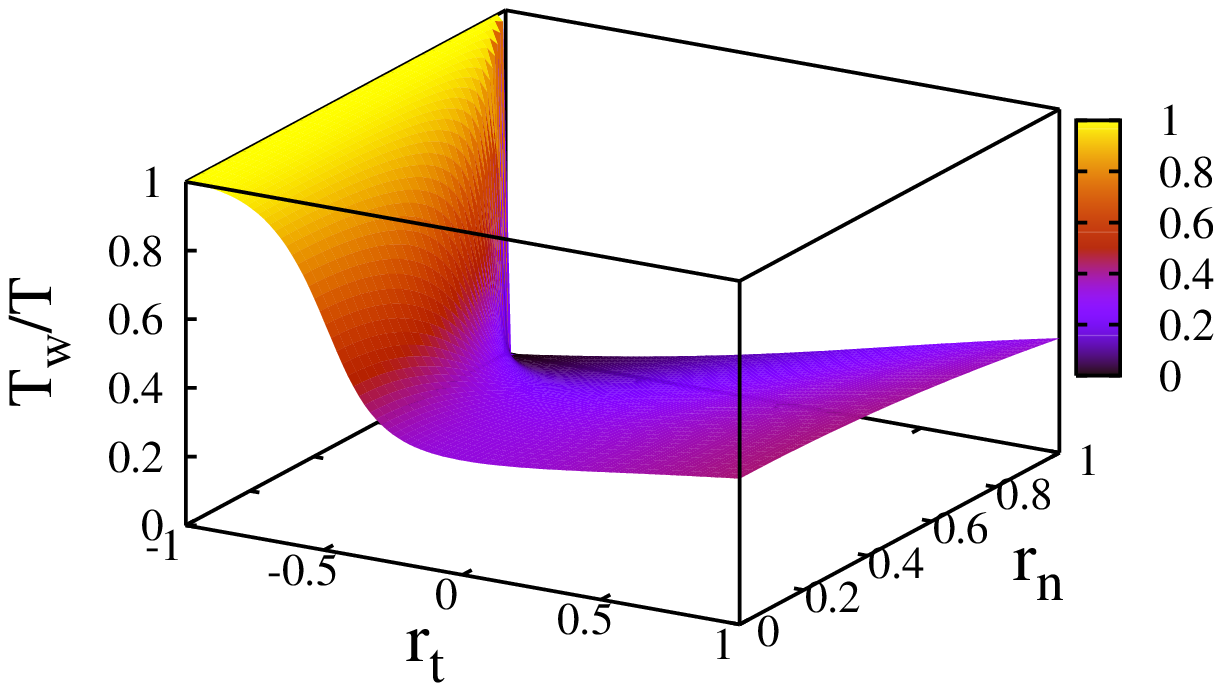}
\label{temperature-ratio}
\caption{The ratio of average rotational energy to total energy as a
  function of the coefficients of normal, $r_n$, and tangential, $r_t$,
  restitution.}
\end{minipage}
\hfill
\begin{minipage}{0.47\textwidth}
\includegraphics*[width=\textwidth]{fig9.eps}
\caption{The scaling function underlying the energy distribution
 (solid line). The distribution was obtained using a Monte Carlo
simulation with $N=10^7$ particles. A dashed line with the slope
predicted by the theory is also shown for reference.\label{energy-free}}
\end{minipage}
\end{figure}

We solve this eigenvalue equation by performing a Monte Carlo
simulation of the same angular process as described by
\eqref{angular} but with a different annihilation rate
$\beta_0=1-\lambda(\nu-1)$. We compare the angle distribution
predicted by \eqref{g-eq-1} with the behavior of the energetic
particles in the freely cooling gas.

The numerical simulations of the inelastic collision process confirm
the theoretical predictions. First, the energy distribution is
self-similar as in \eqref{scaling} and the characteristic scale is
proportional to the temperature. Second, the distribution of the
total energy has a power-law tail, as displayed in
Fig.~\ref{energy-free} and the exponent $\nu$ is very close to the
theoretical prediction (numerical simulations of the collision
process gives $\nu=2.98\pm 0.05$ while the eigenvalue equation
yields $\nu=2.92\pm 0.05$).

\begin{figure}[t]
\begin{minipage}{0.47\textwidth}
\includegraphics*[width=\textwidth]{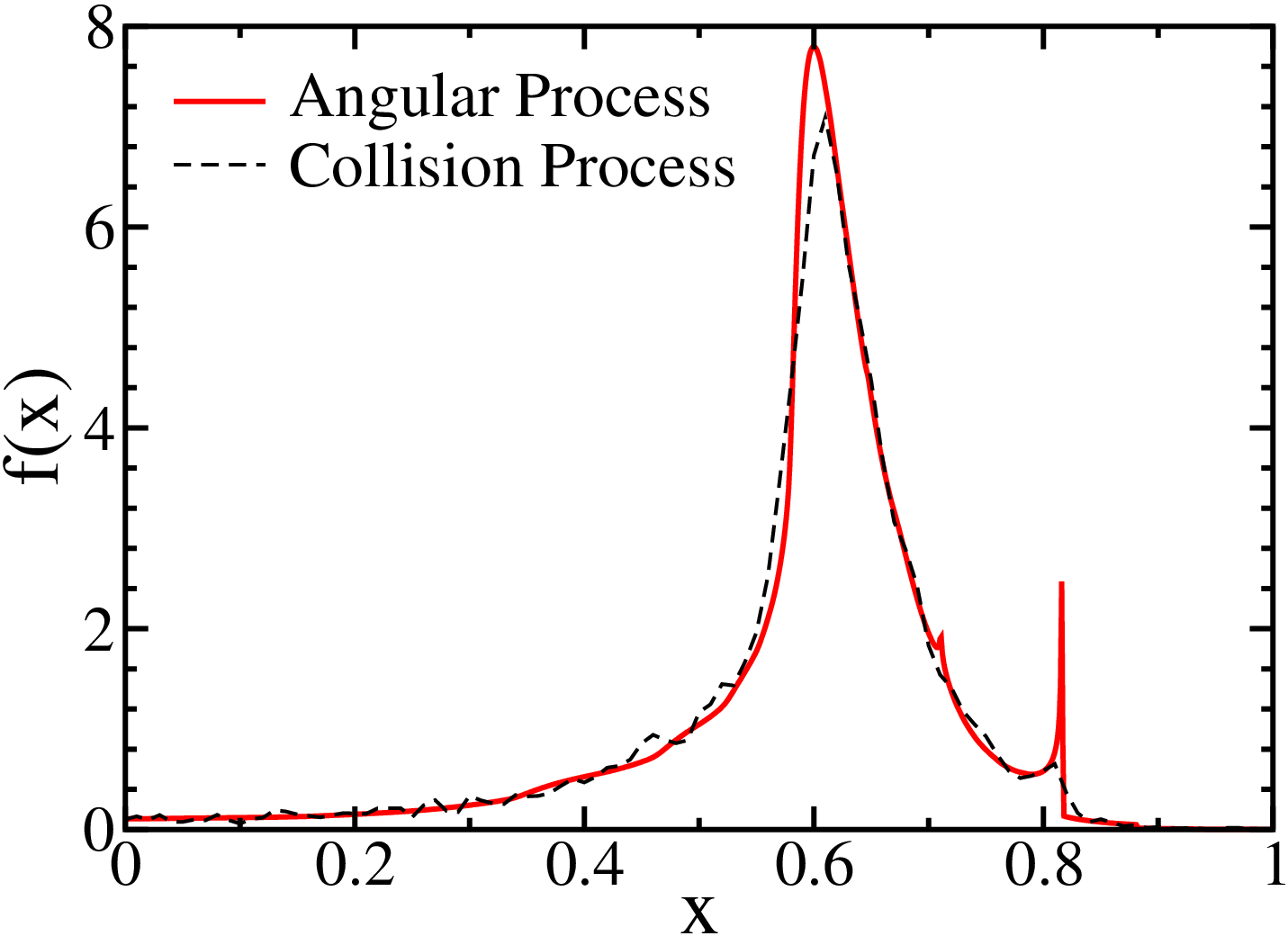}
\caption{The angle distribution of the energetic particles. Shown are
results for the collision process (solid line) and for the angular
process (dashed line).
\label{angle-free}}
\end{minipage}
\hfill
\begin{minipage}{0.47\textwidth}
\includegraphics*[width=\textwidth]{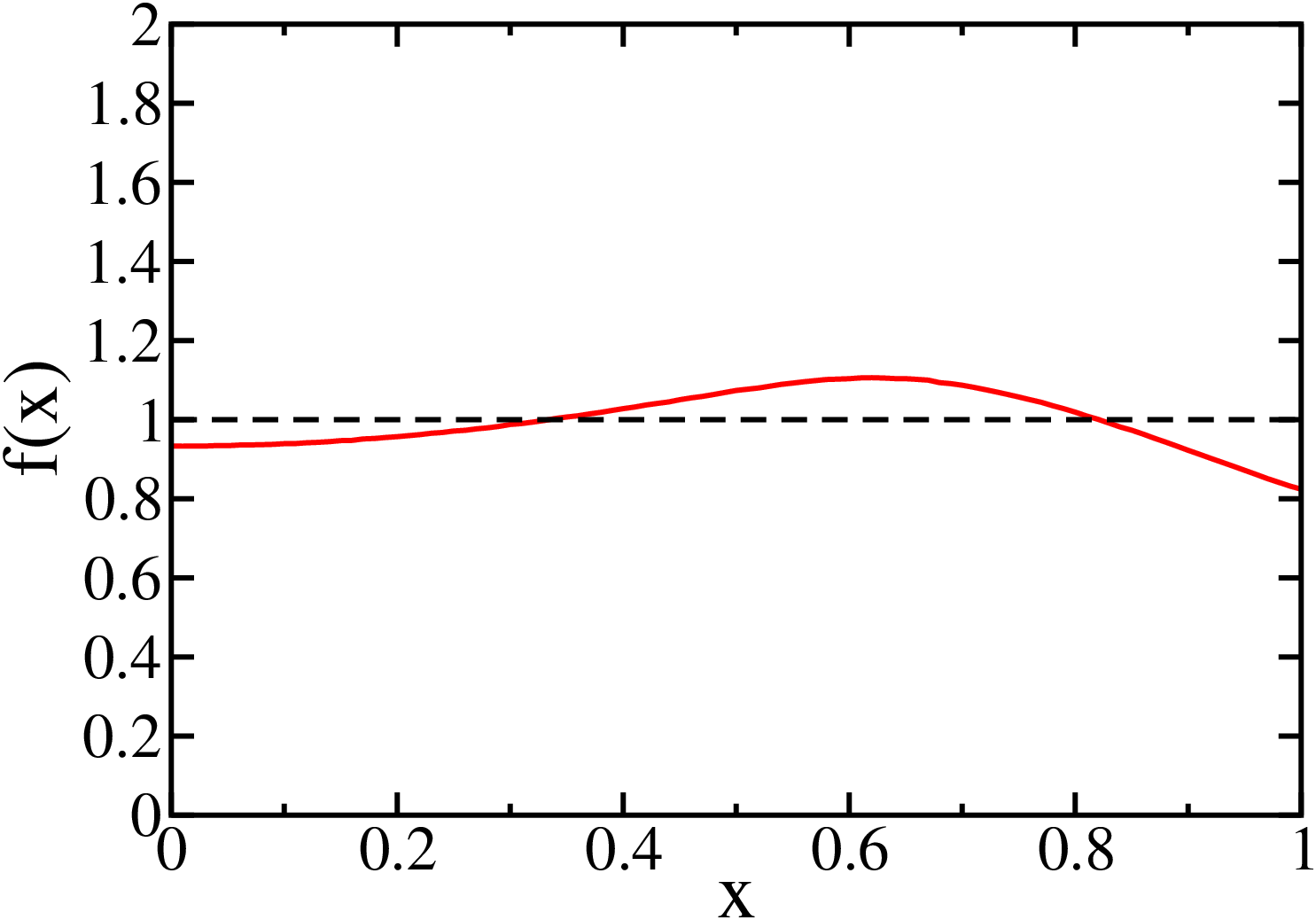}
\caption{The angle distribution of all particles for a freely
cooling gas (solid line). Also shown for reference is the uniform
equilibrium distribution.\label{angle-free-all}}
\end{minipage}
\end{figure}

The angle distribution deviates even more strongly from the uniform
distribution with a very pronounced peak (see Fig.~\ref{angle-free})
because the dynamics are purely collisional. The singularities are
weaker although the one at $x_1$ given by \eqref{special1} is clear.
The agreement between the solution of the angular process and the
Monte Carlo simulations is slightly worse than for driven systems
because the statistics become prohibitive: now it is necessary to
probe the most energetic out of roughly $10^6$ particles to obtain
the asymptotic angle distribution! The sharper power-law decay is
responsible for this three order of magnitude increase: the
cumulative distribution of total energy decays according to $\int_E
dE' p(E')\sim E^{-\mu}$ with $\mu=\nu-1$ about three times larger
than before. Finally, the angle distribution of all particles
deviates only slightly from a uniform distribution (see
Fig.~\ref{angle-free-all}). We conclude that the behavior of the
freely cooling gas is qualitatively similar to that found in driven
steady-states.

\section{Conclusions and Outlook}

The complete description of granular media with translational and
rotational degrees of freedom requires the full bivariate distribution
of energies. It is not sufficient to consider only the average kinetic
energy of translations and rotations. Instead the full bivariate
distribution is highly nontrivial. We have shown that in the limit of
large particle energy, this distribution obeys a linear equation. Its
solution can be written as a product of two distributions, one for the
total energy, $E=E_v+E_w$, and one for the variable $x=\sqrt{E_w/E}$,
which captures the partition of the total energy between rotational
and translational motion. The distribution of the total energy decays
algebraically and the characteristic exponent depends on the collision
parameters and the moment of inertia. The variable $x$ is not
uniformly distributed as in equilibrium. Instead the distribution
$f(x)$ is not analytic and displays a series of singularities of
varying strengths. Remarkably, there are special preferred ratios of
rotational-to-total energy. This violation of energy equipartition
among different degrees of freedom is a direct consequence of the
energy dissipation. The total energy and the variable $x$ are
correlated in general with the deviations from equilibrium increasing
with energy. These two variable become uncorrelated only at extremely
high-energies.

We have studied both, the system which is driven at extremely high
energies and displays a stationary energy cascade on energy scales
below the driving one, and a freely cooling gas. In the latter gas the
bivariate energy distribution is time dependent, reflecting the
overall decrease of energy. Nevertheless, scaling the total energy
with temperature, one finds a self-similar form for the distribution,
which again factorizes in the high-energy limit. As in the driven
system, the distribution of the total energy decays as a power law
with, however, different exponents for the driven and the free cooling
system. The angular distribution deviates even more from the uniform
(equipartition) one in the cooling system.

It should be straightforward to extend these results to three
dimensions where the angular process takes place in three
dimensions. In the limit of high energies one would again expect a
limiting distribution for the partition angle $x=\sqrt{E_w/E}$.
Another possible extension refers to a more realistic law of friction,
including Coulomb friction \cite{Walton,flca}. Finally, it would be of
interest to extend the analysis to other systems, where equipartition
is violated. An example is a binary mixture, where the energy is
shared unequally between the two components.

\acknowledgments We thank the Kavli Institute for Theoretical Physics
in University of California, Santa Barbara where this work was
initiated. We acknowledge financial support from DOE grant
DE-AC52-06NA25396.

\appendix

\section{The Collision rules}

The total linear momentum ${\bf v}'_a+{\bf v}'_b={\bf v}_a+{\bf
v}_b$ is conserved in the collision. The angular momenta of the two
particles with respect to the point of contact, ${\bm \omega}'_i$,
are given by
\begin{subequations}
\label{omega-def}
\begin{align}
I{\bm \omega}_a&=I{\bf w}_a+m\,R\,\hat{\bf n}\times{\bf v}_a\\
I{\bm \omega}_b&=I{\bf w}_b-m\,R\,\hat{\bf n}\times{\bf v}_b.
\end{align}
\end{subequations}
These are conserved, ${\bm \omega}'_i={\bm \omega}_i$ with $i=a,b$,
because there is no torque at the point of contact. In inelastic
collisions, the normal and tangential components of the relative
velocity at the point of contact obey the collision law (\ref{law})
where ${\bf U}={\bf V}+R\,\hat{\bf n}\times{\bf W}$.

It is convenient to introduce the momentum transfer ${\bm \delta}$,
defined as follows: ${\bf v}'_a={\bf v}_a-{\bm \delta}$ and ${\bf
v}'_b={\bf v}_b+{\bm \delta}$.  Conservation of the angular velocity
with respect to the point of contact and Eq.~\eqref{omega-def} gives
${\bf w}'_i={\bf w}_i+\frac{1}{qR}\,\hat{\bf n}\times{\bm \delta}$.
In terms of ${\bm \delta}$, the difference in velocity at the point of
contact is \hbox{${\bf U}'={\bf U}-2{\bm\delta}+\frac{2}{q}\,\hat{\bf
n}\times\hat{\bf n}\times{\bm \delta}$}. Substituting this expression
into the collision laws \eqref{law}, the normal and the tangential
components of ${\bm \delta}$ are simply
\begin{subequations}
\begin{align}
{\bm \delta}\cdot \hat{\bf n}&=\eta_n\,{\bf U}\cdot \hat{\bf n}\\
{\bm \delta}\times \hat{\bf n}&=\eta_t\,{\bf U}\times \hat{\bf n}.
\end{align}
\end{subequations}
Consequently, the momentum transfer is ${\bm \delta}=\eta_n {\bf
U}\cdot\hat{\bf n}\,\hat{\bf n}+ \eta_t({\bf U}-{\bf U}\cdot\hat{\bf
n}\,\hat{\bf n})$ or explicitly,
\begin{equation}
\label{delta} {\bm \delta}=
\eta_n\,{\bf V}\cdot\hat{\bf n}\,\hat{\bf n}+
\eta_t\left({\bf V}-{\bf V}\cdot\hat{\bf n}\,\hat{\bf n}\right)
+\eta_t \,R\,\hat{\bf n}\times{\bm W}
\end{equation}
We now have the explicit collision rules \eqref{rule}.

\section{Particle Number Conservation}
\label{stationary}

In this appendix, we verify that the stationary solution is consistent
with particle number conservation. Maxwell Molecules are considered
for simplicity. It is straightforward to generalize this calculation to
all $\gamma$ and to free cooling.

Our starting point is Eq.~\eqref{linear}, specialized to Maxwell
molecules, i.e. $\gamma=0$,
\begin{equation}
\label{linearA1}
\frac{\partial P(E,\Omega)}{\partial t}\!=\!\int d\Omega_0
\left[\frac{1}{\alpha_1}\!P\left(\!\frac{E}{\alpha_1},\Omega_0\!\right)
\delta(\Omega-\Omega_1)+\frac{1}{\alpha_2}\!
      P\left(\!\frac{E}{\alpha_2},\Omega_0\!\right)
\delta(\Omega-\Omega_2)-P(E,\Omega_0)\delta(\Omega-\Omega_0)\right].
\end{equation}
As a first step we integrate this equation over the solid angle
\begin{equation}
\label{linearA2}
\frac{\partial p(E)}{\partial t}\!=\!\int d\Omega_0
\left[\frac{1}{\alpha_1}\!P\left(\!\frac{E}{\alpha_1},\Omega_0\!\right)
+\frac{1}{\alpha_2}\!P\left(\!\frac{E}{\alpha_2},\Omega_0\!\right)-
P(E,\Omega_0)\right].
\end{equation}
The power-law behavior \eqref{powerlaw} typically holds in a
restricted energy range, $E_{l}\leq E \leq E_{u}$, where $E_l$ and
$E_u$ are upper and lower cutoffs.  In the driven case, the upper
cutoff is set by the energy injection scale.  Let
$N=\int_{E_{l}}^{E_{u}} dE \, p(E)$ be the total number of particles
in this range. With the powerlaw decay \eqref{powerlaw}, then 
\begin{equation}
N\sim \frac{1}{\nu-1}\left(E_{l}^{1-\nu}-E_{u}^{1-\nu}\right).
\end{equation}
To evaluate this time evolution of $N$, we substitute the product form
\eqref{product} into \eqref{linearA2} and integrate over the energies
in the aforementioned power-law range,
\begin{equation}
\label{linearA3}
\frac{\partial N}{\partial t}=N
\int d\Omega_0\, g(\Omega_0)
\left[\alpha_1^{\nu-1}+\alpha_2^{\nu-1}-1\right].
\end{equation}
Using Eq.~\eqref{balance}, we confirm that the total number of
particles is conserved, $\partial N/\partial t=0$.

\section{The matrix coefficients}

\label{coefficents} In an inelastic collision, the translational
energy loss is $\Delta E_v=E_v-E_v'$ with
$E_v=\frac{1}{2}(v_a^2+v_b^2)$ and similarly, the rotational energy
loss is $\Delta E_w=E_w-E_w'$ with $E_w=\frac{1}{2}q(w_a^2+w_b^2)$.
We can conveniently calculate these quantities by using ${\bf
v}'_a={\bf v}_a-{\bm \delta}$, ${\bf v}'_b={\bf v}_b+{\bm \delta}$,
and ${\bf w}'_i={\bf w}_i+(1/qR)\hat{\bf n}\times {\bm \delta}$, and
by expressing the momentum transfer ${\bm\delta}$ using the natural
coordinate system, ${\bm\delta}=\eta_n\,V_n\hat{\bf
n}+\eta_t(V_t-W)\hat{\bf t}$,
\begin{subequations}
\begin{align}
\Delta E_v&=\eta_n(1-\eta_n)V_n^2+\eta_t(1-\eta_t)V_t^2-\eta_t^2W^2+
\eta_t(2\eta_t-1)V_tW \\
\Delta
E_w&=-(\eta_t^2/q)V_t^2+\eta_t(1-\eta_t/q)W^2-\eta_t(1-2\eta_t/q)V_tW.
\end{align}
\end{subequations}
The rate of change of the respective temperatures equals $1/2$ the
average of this quantities, $\frac{d}{dt}T_v=\frac{1}{2}\langle \Delta
E_v\rangle$ and $\frac{d}{dt}T_w=\frac{1}{2}\langle \Delta
E_w\rangle$. This is seen by multiplying \eqref{nonlinear-mm} by
$\frac{1}{2}v^2$ and by integrating over the velocity. The averaging
is with respect to the probability distribution functions of the two
colliding particles. The cross-term vanishes, $\langle V_tW\rangle=0$,
by symmetry. Using $\langle V_n^2\rangle =2\langle
v_n^2\rangle=\langle v^2\rangle=2T_v$ and $\langle W^2\rangle
=2\langle w^2\rangle=4T_w/q$ we obtain the matrix elements
\eqref{lambda}.

\end{document}